\documentstyle[amsfonts,prl,aps,multicol]{revtex}
\input psfig
\pssilent
\begin{document}


\newcommand{\lP}{l_{\rm P}}
\newcommand{\sgn}{{\rm sgn}}

\title{Inflation from Quantum Geometry}
\author{Martin Bojowald\cite{Email}}
\address{Center for Gravitational Physics and Geometry, The
  Pennsylvania State University,\\ 104 Davey Lab, University Park, PA
  16802, USA}

\maketitle

\begin{abstract}
  Quantum geometry predicts that a universe evolves through an
  inflationary phase at small volume before exiting gracefully into a
  standard Friedmann phase. This does not require the introduction of
  additional matter fields with ad hoc potentials; rather, it occurs
  because of a quantum gravity modification of the kinetic part of
  ordinary matter Hamiltonians. An application of the same mechanism
  can explain why the present-day cosmological acceleration is so tiny.
\end{abstract}

\vspace{-5cm} 
\begin{flushright}
CGPG--02/6--2 \\
gr-qc/0206054 \\
\end{flushright}
\vspace{3.4cm}


\begin{multicols}{2}

Inflation (an accelerated phase of the universe, i.e.\ $\ddot{a}>0$
for the scale factor $a$) has been proposed as an elegant solution of
severe problems of the cosmological standard model
\cite{Guth}. However, it suffers from the lack of a natural
explanation from fundamental physics. Usually, a scalar field, the
inflaton, is proposed which requires a special potential and initial
conditions for sufficient inflation. The inflationary phase is usually
deemed to happen after the Planck epoch of the universe, but it has
been speculated that there is a quantum gravity origin.

The first such realization employed effective higher curvature terms
in the gravitational action \cite{Starobinsky} which leads to
deSitter-like solutions, or inflation, even without introducing a
cosmological constant explicitly. However, in this {\em
perturbative\/} higher derivative theory not all solutions are
reliable and non-analytic ones in the perturbation parameter (the
Planck length) have to be excluded \cite{Simon}. In particular, this
has to be done for the deSitter-like solutions since their effective
cosmological constant is given by an inverse power of the Planck
length.

A more recent idea \cite{non-comm} is to use modified dispersion
relations, which are expected to appear in most quantum gravity
theories, for an explanation of inflation. In fact, this is possible
provided that the dispersion relation has a special form with the
momentum decreasing to zero for large energies. Such dispersion
relations have been proposed, but not yet derived from any theory of
quantum gravity. Moreover, in current derivations of dispersion
relations one obtains a power series in the energy which is
inconclusive concerning the high energy behavior.

A lesson is that a non-perturbative approach would produce the most
reliable answer as to whether or not inflation can be derived from
quantum gravity. Currently, the most direct non-perturbative
realization of a quantum theory of gravity is provided by quantum
geometry (also called loop quantum gravity,
\cite{Nonpert,Rov:Loops,ThomasRev}), which is a canonical quantization
of general relativity in terms of Ashtekar's variables. In recent
years, techniques to study its cosmological sector have been developed
\cite{IsoCosmo} which will be used here. We will see that quantum
geometry can naturally lead to inflation at small volume with a
graceful exit.

\paragraph*{Loop quantum cosmology.}

By studying isotropic states of quantum geometry one obtains isotropic
loop quantum cosmology which reproduces the standard Wheeler-DeWitt
approach at large volume \cite{SemiClass}, but leads to significant
deviations at small volume caused by the discreteness of
geometry. Notable consequences are the absence of cosmological
singularities \cite{Sing} and the prediction of `dynamical initial
conditions' for the wave function of a universe \cite{DynIn}. These
already offer solutions to some of the pressing cosmological problems;
here we ask what this implies for the occurrence of inflation.

In what follows we recall the basic formulae necessary for this
investigation.  Basic classical variables are the connection component
$c$ and the isotropic triad component $p$. Because of the two possible
orientations of the triad, $p$ can also take negative values, but
later we will only need positive ones. It is related to the scale
factor $a$ by $|p|=a^2$. For spatially flat models, which will only be
used here, $c$ is proportional to the extrinsic curvature
$k=\frac{1}{2}\dot{a}$. The variables $(k,p)$ form a canonical pair
with $\{k,p\}=\frac{1}{3}\kappa$, $\kappa=8\pi G$ being the
gravitational constant. They appear in the gravitational part of the
Hamiltonian constraint
\begin{equation}\label{Ham}
 {\cal H}=-6\kappa^{-1}k^2\sqrt{p}+{\cal H}_{\phi}
\end{equation}
where ${\cal H}_{\phi}$ is the matter Hamiltonian for a field $\phi$
(this represents a standard matter field and need {\em not\/} be an
inflaton) which for simplicity will be taken to be of the standard
form ${\cal H}_{\phi}=\case{1}{2}a^{-3}p_{\phi}^2+a^3V(\phi)$. Here,
$V(\phi)$ is an arbitrary potential and $p_{\phi}=a^3\dot{\phi}$ the
momentum canonically conjugate to $\phi$. The Hamiltonian constraint
${\cal H}$ generates the dynamics of the theory and gives the
Friedmann equation after using $k=\frac{1}{2}\dot{a}$.

A standard Wheeler-DeWitt quantization would proceed by using wave
functions $\psi(p,\phi)$. Consequently, $\hat{p}$ has a continuous
spectrum containing zero and, after choosing the scale factor as
internal time, the Hamiltonian constraint turns into a second order
differential time evolution equation, the Wheeler-DeWitt equation
\[
 -\case{2}{3}\lP^4\frac{\partial^2}{\partial p^2}(\sqrt{p}\psi)\equiv
 -\case{1}{6}\lP^4a^{-1}\frac{\partial}{\partial a}a^{-1}
 \frac{\partial}{\partial a} (a\psi)=\kappa\hat{{\cal
 H}}_{\phi}\psi\,.
\]
The matter Hamiltonian $\hat{{\cal H}}_{\phi}$ is an unbounded
operator owing to the kinetic term containing an inverse power of the
scale factor. For the matter variables we can choose a standard
quantization which promotes $\phi$ to a multiplication operator and
$p_{\phi}$ to the derivative operator
$-i\hbar\partial/\partial\phi$. 

We will use the same matter field quantization for loop quantum
cosmology. Concerning the geometrical degrees of freedom, however, the
situation is very different. Most notably, the spectrum of geometric
operators like $\hat{a}$ is discrete, which is directly inherited from
the full theory of quantum geometry. This implies that, in a metric
representation, the wave function $s_n(\phi)$ replacing $\psi(p,\phi)$
lives on a discrete space labeled by an integer $n$ representing the
geometry. (Again, the sign of $n$ is the orientation, but later only
positive $n$ will be needed.) The integer $n$ also appears in the
discrete eigenvalues of $\hat{p}$ which have the form
\begin{equation}\label{pn}
 p_n=\case{1}{6}\gamma\lP^2 n
\end{equation}
where a new parameter $\gamma$, the Barbero--Immirzi parameter,
occurs. It is a positive real number which sets the scale for the
discreteness of geometry and thus plays the role of a new
dimensionless fundamental constant which controls the continuum limit
($\gamma\to0$ and $n\to\infty$) just as $\hbar$ controls the classical
limit \cite{SemiClass}. Its value $\gamma=\ln 2/\pi\sqrt{3}\approx
0.13$ has been determined by comparing the black hole entropy obtained
from quantum geometry with the semiclassical result
\cite{ABCK:LoopEntro,IHEntro}. Also the volume operator $\hat{V}$ has
a discrete spectrum with eigenvalues \cite{cosmoII}
\begin{equation}
 V_{\frac{1}{2}(n-1)}=(\case{1}{6}\gamma\lP^2)^{\frac{3}{2}}
 \sqrt{(n-1)n(n+1)}\,.
\end{equation}
Both operators $\hat{p}$ and $\hat{V}$ have eigenvalue zero for
$n=0$. Therefore, the obvious definition of their inverse fails and we
have to face the problem of how to quantize the kinetic part of the
matter Hamiltonian which contains the inverse volume.

\paragraph*{Inverse volume operators.}

\begin{figure}
 \centerline{\psfig{figure=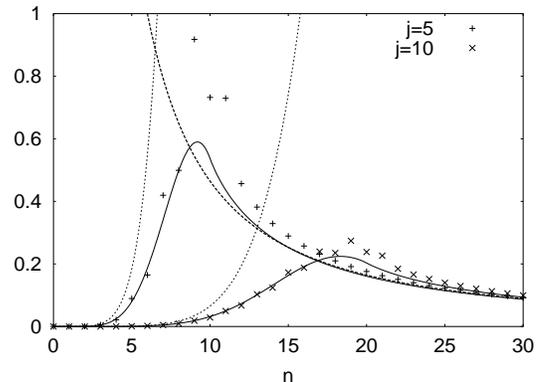,width=3in}}
\caption{Eigenvalues $d_{j,n}$ of the density operator $\hat{d}_j$ (in
units of $(\gamma\lP^2)^{-\frac{3}{2}}$) for two different $j$,
compared to the approximation $p(n/2j)^6V_{\frac{1}{2}(n-1)}^{-1}$
(solid lines) and the classical expectation
$V_{\frac{1}{2}(n-1)}^{-1}$ (thick dashed line). The dotted lines are
the approximations (\ref{djnsmall}) for small $n$.}
\label{dens}
\end{figure}

It turns out \cite{InvScale} that there is a well-defined, finite
quantization of inverse powers of the scale factor in loop quantum
cosmology, which makes use of a general technique of full quantum
geometry \cite{QSDI,QSDV}. This can be interpreted as providing a
natural curvature cut-off and plays a crucial role in the proof of
absence of cosmological singularities \cite{Sing}. Here we illustrate
this technique in the isotropic context: Using the symplectic
structure it is easy to see that the scale factor can be written as
$a=2\kappa^{-1}\{k,V\}$. This implies
\[
 a^{-1}=aV^{-\frac{2}{3}}=2\kappa^{-1}V^{-\frac{2}{3}}\{k,V\}=
 6\kappa^{-1}\{k,V^{\frac{1}{3}}\}
\]
where the negative power of the volume has been absorbed into the
Poisson bracket. This reformulation allows a quantization by
expressing $k$ in terms of a holonomy of the Ashtekar connection
(which is a basic operator in quantum geometry), using the volume
operator and turning the Poisson bracket into a commutator. However,
since the expression for $a^{-1}$ is a rather complicated function of
the basic variables, the final quantization contains quantization
ambiguities of different types. Most of them only affect the lowest
eigenvalues for small $n$ of the order one; but there is one,
resulting from the use of arbitrary representations with label $j$ for
the holonomy, which can also affect higher eigenvalues
\cite{Ambig}. This quantization ambiguity provides the mechanism which
will be exploited here to obtain inflation in an effective classical
description; as with all quantization ambiguities, the final judgement
about their value has to come from observations. We will employ a
quantization $\hat{d}_j$ of the density $d=a^{-3}$ which has
eigenvalues (see Fig.~\ref{dens})
\[
 d_{j,n}= \left[12 (j(j+1)(2j+1)\gamma\lP^2)^{-1}\!\!\sum_{k=-j}^j
 \!\!k\sqrt{V_{\frac{1}{2}(|n+2k|-1)}}\right]^6
\]
labeled by the ambiguity parameter $j$, a half-integer (we use an
operator $\hat{d}_j$ which is the sixth power of the operator
$\hat{s}_j$ of \cite{Ambig}).

\vspace{-4mm}

An approximation, which gets better for large $j$, is obtained as
$d_{j,n}\simeq V_{\frac{1}{2}(n-1)}^{-1}p(n/2j)^6$ where the function
\begin{eqnarray*}
 p(q) &=& \case{8}{77} q^{\frac{1}{4}}
 \left( 7\left((q+1)^{\frac{11}{4}}-
 |q-1|^{\frac{11}{4}}\right)\right.\\
 &&-\left.11q\left( (q+1)^{\frac{7}{4}}- \sgn(q-1)
 |q-1|^{\frac{7}{4}}\right) \right)
\end{eqnarray*}
has been derived in \cite{Ambig}.  This formula has the asymptotic
expansions
\begin{equation}\label{djnsmall}
 d_{j,n}\simeq \case{12^6}{7^6} V_{\frac{1}{2}(n-1)}^{-1}
 (n/2j)^{\frac{15}{2}}
\end{equation}
for $n\ll 2j$ and $d_{j,n}\simeq V_{\frac{1}{2}(n-1)}^{-1}$ for $n\gg
2j$. The approximation (\ref{djnsmall}) shows how the classical
divergence of the inverse volume is truncated by the large power of
$n$. In particular, the dependence on $n$ of the density $d_j$ is very
different at small and large $n$ which, when used in the matter
Hamiltonian, implies severe changes of the cosmological evolution near
the classical singularity. If $j$ is larger than one, we can study
this evolution by an effective matter Hamiltonian which incorporates
these different behaviors. To this end, we write the density as a
function of $a$ instead of $n$ using the relation
$a^2=\frac{1}{6}\gamma\lP^2n$ which follows from (\ref{pn}). This
yields
\begin{equation}\label{dja}
 d_j(a)=a^{-3}p(3a^2/\gamma\lP^2j)^6
\end{equation}
with approximations
\begin{equation}\label{djaappr}
 d_j(a)\simeq \case{12^6}{7^6}(\case{1}{3}\gamma\lP^2j)^{-\frac{15}{2}}a^{12}
\end{equation}
for $a^2\ll\frac{1}{3}\gamma\lP^2j$ and $d_j(a)\simeq a^{-3}$
for $a^2\gg\frac{1}{3}\gamma\lP^2 j$. The approximation for small $a$
disappears in the continuum limit $\gamma\to0$, in which case we only
obtain the classical behavior $a^{-3}$.

\paragraph*{Effective Friedmann equation.}

Since the spectra of geometric operators in loop quantum cosmology are
discrete, the evolution equation in an internal time variable will
also be discrete, i.e.\ a difference rather than a differential
equation \cite{cosmoIV}. Its main part is of the form \cite{IsoCosmo}
\begin{eqnarray}\label{WdW}
&&({\cal D}s)_n(\phi):=(V_{\frac{1}{2}(n+4)}\!-V_{\frac{1}{2}(n+4)-1})
 s_{n+4}(\phi)\nonumber\\ &&-
 2(V_{\frac{1}{2}n}\!-V_{\frac{1}{2}n-1}) s_{n}(\phi)
 +(V_{\frac{1}{2}(n-4)}-V_{\frac{1}{2}(n-4)-1}) s_{n-4}(\phi)\nonumber\\
 &&=-\case{1}{3}\gamma^3\kappa l_{\rm P}^2\, \hat{{\cal
 H}}_{\phi}(n)s_n(\phi)
\end{eqnarray}
where $n$ (analogous to $p$) is the discrete internal time (negative
$n$, which are not being considered here, correspond to time before
the classical singularity \cite{Sing}). If the wave function is not
oscillating at small scales (from $n$ to $n+1$), the difference
operator ${\cal D}$ can be approximated by a differential operator
which turns out to be $3\gamma^{-3}\lP^{-2}{\cal D}\approx
\frac{2}{3}\lP^4\frac{\partial^2}{\partial p^2}\sqrt{p}$, i.e.\ the
operator appearing in the standard Wheeler--DeWitt equation
\cite{SemiClass}. Noting that
$\frac{1}{3}i\lP^2\frac{\partial}{\partial p}$ quantizes
$k=\frac{1}{2}\dot{a}$, the resulting evolution equation corresponds
to the Friedmann equation $\frac{3}{2}\dot{a}^2a=\kappa{\cal
H}_{\phi}(a)$. The matter Hamiltonian, however, does not have the
Wheeler--DeWitt form for all $a$ because we were forced to quantize
the inverse volume to the well-defined operator $\hat{d}_j$ ($j$ can
be regarded as a new quantum number characterizing the matter
component). Therefore, instead of the density $a^{-3}$ we have the
effective density $d_j(a)$ from (\ref{dja}) which is modified at small
$a$. Our Friedmann equation for $H=\dot{a}/a$ then is
\begin{equation}\label{Friedmann}
 H^2=\case{2}{3}\kappa a^{-3}{\cal H}_{\phi}^{(j)}(a)=
 \case{2}{3}\kappa (\case{1}{2}a^{-3}d_j(a)p_{\phi}^2+V(\phi))
\end{equation}
which will be studied in the remainder of this paper.

\paragraph*{Super-inflation.}

\begin{figure}
 \centerline{\psfig{figure=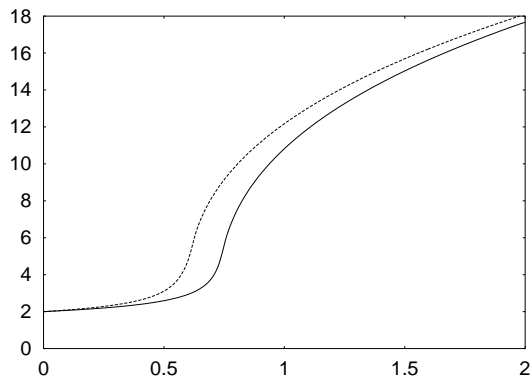,width=3in}}
\caption{Numerical solutions $a(t)$ (in units of $\sqrt{\gamma}\lP$,
with an arbitrary scale for $t$) of the effective Friedmann equation
(\ref{Friedmann}) for $V(\phi)=0$ (solid) and a small quadratic
potential (dashed), both with $j=100$.}
\label{infl}
\end{figure}

To simplify the analysis we first assume that the matter field $\phi$
is a free massless field, i.e.\ $V(\phi)=0$. Then, $p_{\phi}$ is
constant and at large volume we have the Friedmann phase $a(t)\propto
(t-t_0)^{1/3}$. If $a^2$ is small compared to
$\frac{1}{3}\gamma\lP^2j$, however, the density behaves as in
(\ref{djaappr}) which leads to super-inflationary expansion
\cite{GeneralInfl} $a(t)\propto (t_0-t)^{-\frac{2}{9}}=
(t_0-t)^{\frac{2}{3(1+w)}}$ with an equation of state parameter
$w=-4$. ($p_{\phi}$ has to be non-zero, but even a quantum fluctuation
would suffice to get the inflationary growth started.) There is a pole
at $t=t_0$, but it is never reached because the small-$a$
approximation breaks down when $a$ becomes larger. In fact, the
super-inflationary phase automatically ends when the peak of $d_j(a)$
is reached, exhibiting a graceful exit into the Friedmann phase
(Fig.~\ref{infl}). Note also that $w$ is not constant during
inflation; in fact it varies smoothly between $-4$ for small $a$ and
$w=1$ for large $a$. A non-zero potential does not change this
behavior qualitatively; depending on its form, however, it could lead
to a second phase of potential driven inflation. This would occur more
generically than in the standard scenario because the first phase can
drive the scalar field up the potential hill, even if $\phi$ starts at
the value $\phi=0$.

\vspace{-4mm}

\vspace{-4mm}

\begin{figure}
 \centerline{\psfig{figure=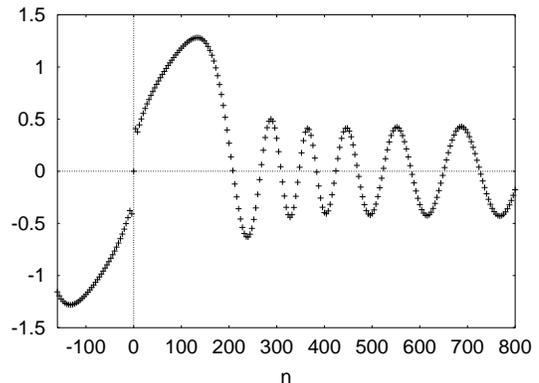,width=3in}}
\caption{An example for the discrete wave function $s_n(\phi)$ with a
free massless scalar field and $j=200$. Negative $n$ represent time
before the classical singularity.}
\label{inflwave}
\end{figure}

An inflationary expansion can also be seen from the wave function
$s_n(\phi)$ (Fig.~\ref{inflwave}; see \cite{Scalar} for details). At
small $n$ the amplitude of the oscillations and the oscillation length
decrease toward larger $n$ (indicating acceleration). After this phase
(for $n>2j=400$ in the figure), the oscillation length increases
(deceleration).

\vspace{-4mm}

\vspace{-4mm}

The appearance of an inflationary phase is a direct consequence of the
modified behavior of the density $d_j(a)$ for
$a^2<\frac{1}{3}\gamma\lP^2j$. This shows that such a phase disappears
in the continuum limit $\gamma\to0$ (which leads to Wheeler--DeWitt
quantum cosmology) or in the classical limit $\lP\to0$. It is only
possible if we incorporate the discreteness which requires $\gamma>0$,
and $\lP>0$.

The coefficient $w$ which determines the rate of super-inflationary
growth is not unique and depends on the quantization choices. Instead
of $\hat{d}_j$ one can, e.g., use an operator
$\hat{d}_{m,j}$ with eigenvalues
\[
 d_{m,j,n}\!=\!\!\left[\!12(j(j+1)(2j+1)\gamma\lP^2)^{-1}\!\!\!\sum_{k=-j}^j
 \!kV_{\frac{1}{2}(|n+2k|-1)}^m\right]^{\frac{3}{2-3m}}
\]
with $0<m<\frac{2}{3}$ whose eigenvalues at large $n$ drop off as
$n^{-\frac{3}{2}}$.  At small $a$ the behavior is $d_{m,j}(a)\propto
a^l$ where the power $l=\frac{6}{2-3m}$ is bounded from below by
$l>3$. This leads always to super-inflationary growth
$a\propto(t_0-t)^{\frac{2}{3-l}}$ with coefficient $w=-l/3$ but can
get arbitrarily close to standard inflation ($w=-1$).
In fact one can also obtain phases of standard or power-law
inflation \cite{PowerLaw}, though in a less natural way (e.g.\ by
using $\hat{d}_{1}\hat{d}_{j,1/3}^{2/3}\hat{V}^{2/3}$ which grows
as $a^3$ for $\gamma\lP^2\ll a^2\ll \gamma\lP^2j$).

\paragraph*{Remarks.}

One may wonder how a quantum gravity effect can influence the universe
expansion at large scales without a violation of the classical
limit. This is possible because the limit $\lP\to0$ must include
$n\to\infty$ in order to keep $a^2=\frac{1}{3}\gamma\lP^2n$
non-zero. For $n\to\infty$ we are always in a Friedmann phase, no
matter how large $j$ is. However, $\lP$ is non-zero and the classical
limit must not be performed completely. This leaves open the
possibility of quantum gravity effects at large volume.

Choosing very large values for the parameter $j$ leads to matter
Hamiltonians with kinetic terms which even today are still growing
with $a$ and so exhibit a super-inflationary phase at the present time
(this matter component would behave as ``phantom matter'' with $w<-1$
\cite{phantom}). While this involves huge $j$ and so is less natural,
it can explain why today's cosmological constant is so small thanks to
a suppression by an inverse power of $j$: If we assume the potential
term to be negligible, i.e.\ $V(\phi)\approx0$ and thus
$p_{\phi}\approx{\rm const.}$, the Friedmann equation
(\ref{Friedmann}) implies
\[
 H^2\propto (a/\sqrt{j}\lP)^9(\sqrt{j}\lP)^{-6}p_{\phi}^2
\]
when the small-$a$ approximation is valid. Here, the first factor is
smaller than one thanks to $a^2<\frac{1}{3}\gamma\lP^2j$ in the
accelerated phase such that $H^2<(\sqrt{j}\lP)^{-6}p_{\phi}^2$ which
is well below the Planck mass for large $j$.

\paragraph*{Conclusions.}

In this paper we presented what we believe is the first direct
derivation of inflation from a candidate for a quantum theory of
gravity. The arguments have been given in the context of isotropic
models, but we only used techniques and generic properties which apply
in full quantum geometry. Therefore, {\em inflation with a graceful
exit into a standard Friedmann phase can be regarded as a natural
prediction of quantum geometry}. In fact, the inverse volume increases
in a neighborhood of $a=0$ which leads to an accelerated expansion
after evolving through $a=0$ (which is not a singularity in loop
quantum cosmology \cite{Sing}). The modified behavior of the inverse
volume is always present for small values of $a$ below the Planck
scale (even for the minimal $j=\frac{1}{2}$). However, in this regime
the classical space-time picture breaks down and the evolution cannot
be described simply by a scale factor. This region has to be better
understood before one can find a reliable estimate of the scale factor
at the start of inflation.  While it is clear that inflation ends when
the scale factor reaches a value $a^2\approx\frac{1}{3}\gamma\lP^2j$,
its beginning lies in the Planck regime and can in principle be
arbitrarily close to zero. Its value determines the number of
$e$-foldings of the inflationary phase.  The question of how inflation
can be described in the quantum regime and its implications for
fluctuations deserve further study.

The same effect can explain the present-day cosmological
acceleration. While this appears less natural, it implies that if the
mechanism proposed here is responsible, today's cosmological
acceleration {\em has to be small}.

\paragraph*{Acknowledgements.}

The author is grateful to A.\ Ashtekar for a critical reading of the
manuscript.  This work was supported in part by NSF grant PHY00-90091
and the Eberly research funds of Penn State.

\end{multicols}
\end{document}